\documentclass[twocolumn,prl,showpacs,superscriptaddress]{revtex4}
\usepackage{amssymb}
\usepackage{amsmath}
\usepackage{graphicx}
\usepackage{subfigure}
\usepackage{natbib}
\usepackage{epsfig}
\usepackage{amsfonts}
\usepackage{mathrsfs}
\usepackage{ulem}
\usepackage{color}
\usepackage[toc,page,title,titletoc,header]{appendix}

\begin{document}

\title{Fractional Topological States in Quantum Spin Chains with Periodical Modulation}

\author{Haiping Hu}
\affiliation{Beijing National Laboratory for Condensed Matter
Physics, Institute of Physics, Chinese Academy of Sciences,
Beijing 100190, China}
\author{Huaiming Guo}
\affiliation{Department of physics, Beihang University, Beijing 100191, China}
\author{Shu Chen}
\thanks{Corresponding author, schen@aphy.iphy.ac.cn}
\affiliation{Beijing National
Laboratory for Condensed Matter Physics, Institute of Physics,
Chinese Academy of Sciences, Beijing 100190, China}
\affiliation{Collaborative Innovation Center of Quantum Matter, Beijing, China}
\begin{abstract}
We report the findings of fractional topological states in one-dimensional periodically modulated quantum spin chains with up to third neighbor interactions. By exact numerical studies, we demonstrate the existence of topologically nontrivial degenerate ground states at some specific magnetizations, which can be characterized by the nonzero-integer total Chern numbers of the degenerate ground states and the emergence of nontrivial edge states under open boundary conditions. We find that the low-energy excitations obey bosonic $\nu=1/2$ fractional statistics for the spin-$1/2$ system and $\nu=1$ non-Abelian statistics for the spin-$1$ system, respectively. The discovered fractional quantum states provide another route to the theoretical exploration of fractional quantum states in correlated spin systems.
\end{abstract}

\maketitle

{\it Introduction.-} Exploring strongly correlated topological states has become an important and active field of research in condensed matter physics \cite{WangZhong,Gurarie,Kitaev,Senthil,Wen2011,Pollmann2} since the theoretical prediction and experimental observation of topological insulators \cite{review}. The most famous strongly correlated topological state, fractional quantum Hall (FQH) state, clearly demonstrates the central role of the interplay of long-range interaction and Landau levels in low dimensional quantum systems. Recently, many efforts \cite{dnsheng,yfwang1,yfwang2,regnault,bernervig,ylwu,LiuZhao} have been devoted to searching for robust fermionic and bosonic fractional topological phases without Landau levels in topological flat bands to mimic FQH effect on two-dimensional (2D) lattices. The so-called fractional Chern insulators (FCIs) \cite{fci}, which support exotic fractional statistics, may provide a promising platform for implementing topological quantum computation \cite{quantumcomputation}. Besides, topologically nontrivial phases in spin systems have also attracted intensive studies in the past years \cite{Greiter,Yao,Chisnell,Wen2011,Pollmann2}. Particular attention has been devoted to the one-dimensional (1D) quantum spin systems \cite{Wen2011,Pollmann2} as even simple spin chain systems, e.g., the spin-1 Haldane chain \cite{haldane}, can exhibit rich topologically nontrivial phenomena. Taking long-range interaction into account, the famous Haldane-Shastry \cite{haldanemodel,shastrymodel} model and its extension to higher symmetry can host Abelian and non-Abelian statistics regarding the spinon excitations with topological degeneracy encoded in the fractional momentum spacings for the spinons.

By introducing an additional virtual dimension, recently it was proposed that one can generate topologically nontrivial states in the higher-dimensional parameter space \cite{lang,kraus,LiLH,kraus2,Prodan,zhangfan,ortix,Mei}. Various schemes of dimensional extension have been applied to simulate the four-dimensional Hall effect by using cold atoms in a three-dimensional optical lattice with a synthetic extra dimension \cite{4dQHE,kraus2} and the virtual $(1+1)$-dimensional Chern insulator in periodically (or quasi-periodically) modulated 1D systems \cite{lang,kraus,ortix,Prodan,LiLH,Mei}. Motivated by these developments, the periodically modulated spin chains have been rescrutinized from a topological point of view \cite{hu1,hu2}, which unveils magnetization plateau states \cite{oshikawa} in these periodic spin systems having topologically nontrivial properties.
It is well known that FQH state is a collective behavior induced by the long-range Coulomb interaction with the landau level fractionally filled, and similarly FCIs are generated by introducing short-range interactions on fractional filled topological flat bands. Simulating FQH effect  with dimension extension is also studied \cite{xuzhihao,zhai} and fractional topological state is proposed in 1D optical superlattices \cite{xuzhihao}. Stimulated by these observations, in this work we explore fractional topological states, with nontrivial low-energy excitations fulfilling fractional statistics, in periodically modulated quantum spin-$S$ chains including short-range exchange interactions. Different from the fermionic systems, the quantum spin chains have no simple noninteracting analogy except for the spin-$1/2$ chain with only nearest-neighbor (NN) exchange interactions, making it challenging to search and stabilize exotic spin states.

{\it Model.-} We consider the period-$q$ quantum spin-$S$ chains with next-nearest-neighbor (NNN) and next-next-nearest-neighbor (NNNN) exchange interactions, described by the following Hamiltonian:
\begin{eqnarray}
H=\sum_i J_i(S_i^x S_{i+1}^x+S_i^y S_{i+1}^y+g S_i^z S_{i+1}^z)\\\notag
+U\sum_i \vec{S}_i\cdot\vec{S}_{i+2}+\kappa U\sum_i \vec{S}_i\cdot\vec{S}_{i+3}
\end{eqnarray}
with $J_i=J_{q+i}$. Here $g$ represents the anisotropic parameter of the NN exchange interactions. The NNN and NNNN exchange interaction strengths are $U$ and $\kappa U$, respectively. In this work, the periodically modulated exchange coupling is taken as
\begin{equation}
J_i = J [ 1- \lambda \cos(\frac{2\pi}{q} i + \delta)],  \label{Ji}
\end{equation}
which describes cosine modulations with modulation amplitude $\lambda$,  period $q$ and a phase parameter $\delta$. For convenience, we shall take $J=1$ as the unit of energy and focus on the anti-ferromagnetic (AFM) couplings with $J>0$ and $|\lambda|<1$.  To be reliable, NN interaction and NNN interaction strengths are set as $U<1$ and $\kappa<1$. For simplicity, only the $q=3$ case will be considered.

As the $z$-component of total spin $S_z=\sum_i S_i^z$ is a conserved quantity, next we shall explore the fractional topological states in both the spin-1/2 and spin-1 systems with given magnetizations, where the magnetization is defined as $m_z = S_z/L$ with the chain length $L$. As will be demonstrated in the present work, fractional topological states can only be found in systems with some particular values of magnetization. The state with a given magnetization can be always achieved by tuning the strength of an applied magnetic field.

{\it Topologically nontrivial ground state for the spin-$1/2$ system.-} We first study the period-$3$ quantum spin-$1/2$ chain with NNN and NNNN exchange interactions and consider the system with magnetization $m_z=\pm 1 / 3$, for which we find the existence of topologically nontrivial double-degenerate ground states and low-energy excitations fulfilling fractional statistics in a wide parameter regime. To be specific, we perform exact diagonalization (ED) and density matrix renormalization group (DMRG) calculations on the low-energy spectrum and see the properties of the ground states for the specific magnetization $m_z= - 1 / 3$. The chain with length $L$ is divided into $L/3$ unit cells with each unit cell contains three distinguished sites. In a special limit with $U=0$ and $g=0$, the system with $m_z= - 1 / 3$ corresponds to a spinless fermion system with the lowest band being half-filled. However, such a mapping no longer holds true in the presence of  the NNN and NNNN exchange interactions.
\begin{figure}
\includegraphics[width=3.7in,height=2.9in]{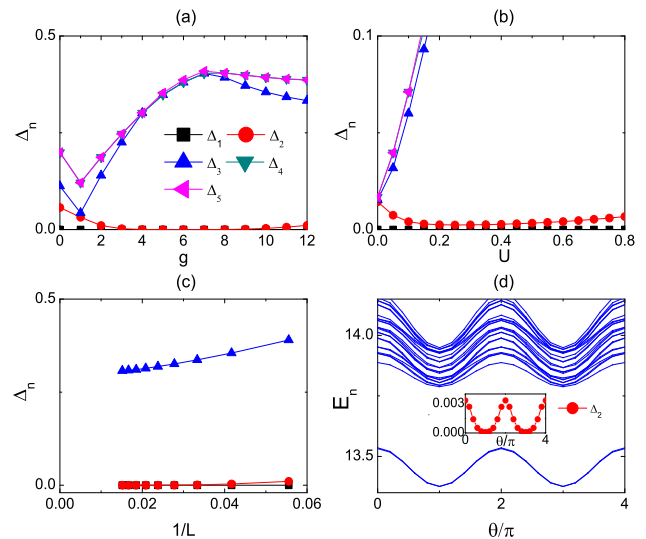}
\caption{(Color online) (a) Excitation energies $\Delta_n$ with respect to the anisotropic parameter $g$ with $U=0.5$. (b) $\Delta_n$ versus $U$ for the system with $g=10$. (c) Finite size analysis of the lowest three excitation energies for systems with $g=10$ and $U=0.5$. (d) Low-energy spectrum with respect to the twist phase $\theta$ as a probe of topological stability for the system with $g=10$, $U=0.5$ and $L=24$. The inset shows the ground states cross each other two times after period-$4\pi$. The other parameters are $\lambda=0.4$, $\kappa=0.4$ and $\delta=0$. }
\end{figure}
Denote the ground state energy as $E_1$ and define $\Delta_n=E_n-E_1$, where $E_n$ is the corresponding energy of the $n$-th many-body state $\Psi_n$. In Fig.1(a), we show the dependence of the excitation energies with respect to $g$ by fixing $U=0.5$ for the system with $m_z= - 1 / 3$ and $L=24$.  When the anisotropic parameter $g$ is small, the ground state is unique. As $g$ increases, the two lowest states tend to be degenerate and there exist a large excitation gap above the 2-fold nearly degenerate ground states. To see the effect of $U$, we show the excitation energies versus $U$ in Fig.1(b) by fixing $g=10$. To rule out the finite size effect, we do finite size analysis of the excitation energies for different chain lengths till $L=66$ by DMRG for systems with $g=10$ and $U=0.5$. As indicated in Fig.1(c), the ground state is $2$-fold degenerate and the large excitation gap between the 2rd and the 3th state still exists in thermodynamical limit. The topological stability of the ground state is demonstrated by imposing twist boundary conditions \cite{twist,hirano,GuoHM} $\psi(j+L)=e^{i\theta}\psi(j)$ on the spin chains, where $j$ denotes an arbitrary site and $\theta$ the twist phase. As shown in Fig.1(d), the 2-fold degenerate ground states evolve into each other and after period-$4\pi$ pumping of the twist phase, the ground state manifold returns to the original configuration as the two ground states must cross each other two times (the inset of Fig.1(d)).

Next we demonstrate the non-trivial topological property of the ground state manifold.
Usually, the existence of edge states are thought to be a hallmark of non-trivial bulk topology. We consider the evolution of edge states with respect to the phase parameter $\delta$. In Figure.2(a) and (b) the excitation energies $\Delta_n$ for system with $L=24$ under periodic boundary conditions (PBC) and open boundary conditions (OBC) are shown respectively. Under PBC, with the evolution of the phase $\delta$, the nearly degenerate two lowest states form a  ground-state manifold, which is well separated from the higher states by a gap. The energy spectrums display quite different behaviors under OBC. The two nearly degenerate ground states for PBC are lifted. One of them connects to the higher spectrum. Once the touching happens, the spin distribution $\langle S_i^z\rangle=\langle\Psi_2|S_i^z|\Psi_2\rangle$ of the in-gap states changes its direction as illustrated in Fig.2(c). For $\delta=\frac{\pi}{5}$, the in-gap state is pinned on the left side, while for $\delta=\frac{9 \pi}{5}$ it is pinned on the right side.

The topological properties of the system under PBC can be characterized by Chern numbers defined in the 2D extended parameter space spanned by $(\theta,\delta)$ \cite{lang,kraus,ortix,Prodan,LiLH,Mei}.
For the many-body state $\psi$, the Chern number is the integral of Berry curvature in parameter space $(\theta,\delta)$:
\begin{eqnarray}
C=\frac{1}{2\pi}\int d\theta d\delta F(\theta,\delta),
\end{eqnarray}
where the Berry curvature \cite{thouless,niuqian} is
\begin{eqnarray}
F(\theta,\delta)=Im(\langle\frac{\partial\psi}{\partial\delta}|\frac{\partial\psi}{\partial\theta}\rangle-
\langle\frac{\partial\psi}{\partial\theta}|\frac{\partial\psi}{\partial\delta}\rangle) .
\end{eqnarray}
Numerically, the calculation of Chern numbers needs division of 2D continuous parameter space into discrete manifold \cite{chern}. We have analyzed different partitions of the parameter space $(\theta,\delta)\in [0,2\pi]\times[0,2\pi]$. For the $2$-fold ground states, while Chern number for each state depends on the partition, their sum is exactly $1$, which indicates the total Chern number being an integer.
\begin{figure}
\includegraphics[width=3.7in,height=3.2in]{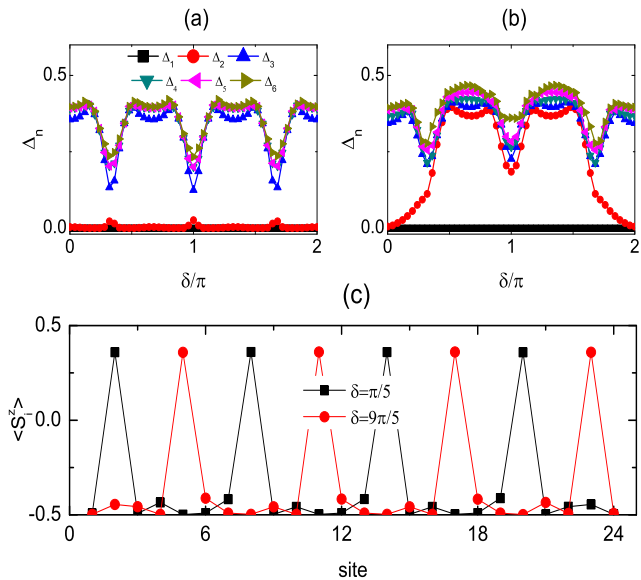}
\caption{(Color online) Excitation energies $\Delta_n$ with respect to the phase parameter $\delta$ under (a) PBC and (b) OBC. (c) The spin distribution for $\delta=\pi/5$ and $\delta=9\pi/5$ for the second state under OBC. The parameters are $U=0.5$, $\lambda=0.4$, $\kappa=0.4$, $g=10$ and $L=24$.}
\end{figure}

{\it Fractional statistics.-} The above observed degenerate ground state manifold and non-trivial topological properties in periodically modulated quantum spin chains with NNN and NNNN interactions resemble the $\nu=1/2$ bosonic FQH state. To remove out any doubt about the fractional nature of this topological quantum state, we investigate the low-energy excitations near the considered magnetization  $m_z=-1/3$. For the spin-$1/2$ chain with length $L$, we consider the low-energy excitations with magnetization $m_z=-1/3 - 1/L$. The reason for this deviation will be explained below. In Fig.3, we calculate the low-energy spectrums for two systems with different lengths $L=18$ and $L=24$. Likely, the topological stability of these low-energy excitations is investigated by imposing twist boundary conditions. For $L=18$, there exists $9$ lower energy states separated by a large excitation gap from the higher excitation states. While for $L=24$, the number of states below the excitation gap is $16$.
\begin{figure}
\includegraphics[width=3.7in,height=3.3in]{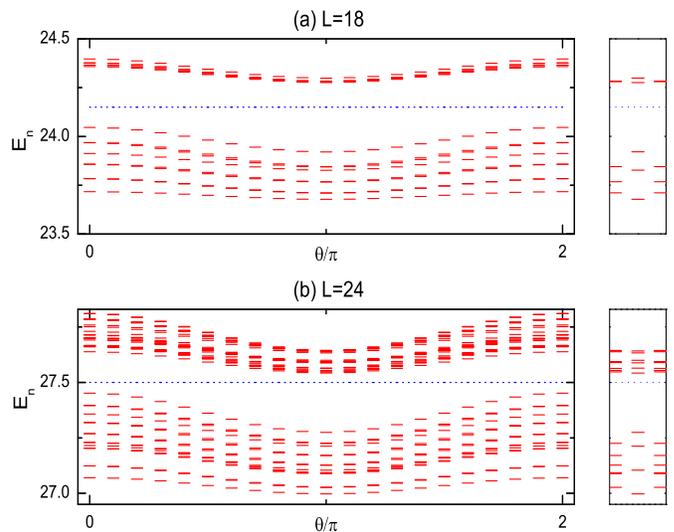}
\caption{(Color online ) Low-energy spectrum $E_n$ under twist boundary conditions for systems with length (a) $L=18$ and (b) $L=24$. There are $9$ and $16$ lower energy states, respectively, which are separated by obvious gaps from the higher excite states. The parameters are $\lambda=0.4$, $U=0.5$, $\kappa=0.4$, $g=10$ and $\delta=0$. The right inserts show spectrums for systems of $\theta=\pi$, in which the degenerate levels are explicitly displayed.}
\end{figure}

The results can be explained by postulating a generalized Pauli principle \cite{regnault} with no more than $1$ particle in $2$ consecutive orbitals after noticing the equivalence between the quantum spin-$1/2$ representation and the hard-core boson representation by the Jordan-Wigner transformation. This transformation maps the magnetization operator to the particle number operator through $S_i^z=b_i^{\dag}b_i-1/2$, and thus the total boson number is given by $N_b = (m_z + 1/2) L $. Take $L=24$ as an example. For the period $q=3$, the number of unit cells $8$ is the orbital number which can be occupied by hard-core bosons. For the system with magnetization $m_z=-1/3$, the problem reduces to how $4$ hard-core bosons fill up $8$ orbitals. Obviously, only the following two configuration: $(10101010)$ and $(01010101)$ are allowed by the (1,2)-admissible counting rule which gives 2-fold degenerate ground states. For the system with magnetization $m_z=-1/3 - 1/L$ which is slightly deviated from $m_z=-1/3$, we need to count how many ways one can remove one boson from the above two ground-state configurations. Only two kinds of configurations are possible, i.e., $(10101000)$ and $(10100100)$ regardless of lattice translation. Finally there are total $16$ lower energy states allowed by the (1,2)-admissible counting rule after $8$ translations of the above $2$ states, which coincides with our numerical results.

{\it Spin-1 system.-}
Next we investigate the periodic quantum spin-$1$ chains with NNN and NNNN interactions.
Generally speaking, the integer quantum spin chains exhibit quite different behaviors from half-integer systems based on low-energy effective field theory \cite{haldane}, which leads to gapped excitation for the former and gapless excitations for the latter.
For the periodically modulated spin-$1$ chain with $q=3$ including NNN and NNNN interactions, we find the emergence of fractional non-Abelian states for magnetization $m_z=(1/q-1)$ with appropriate parameters. Fig.4 shows our main results. First, tuning the anisotropy $g$ ($g\approx 14$) leads to three nearly degenerate ground states which are separated from the 4-th state by a large excitation gap as shown in Fig.4(a). The ground degeneracy is equal to that of the Moore-Read Pfaffian state \cite{mooreread} at filling $\nu=1$. The degeneracy is sensitive to $U$ as shown in Fig.4(b). What we really care about is the existence of fractional excitations. For the $S=1$ chain with periodical modulation, we consider the low-energy spectrum for magnetization $m_z= - 2/3 -1/L$ which is a little deviation from $m_z= - 2/3$ as will be explained below. The topological stability of the low energy spectrums with respect to the twist phase $\theta$ are shown in Fig.4(c)-(d) for chain lengths $L=12$ and $L=15$, respectively. For the former case, there exist $8$ lower energy states which are quite stable with respect to the twist phase. While for $L=15$, we observe $15$ states which are separated from higher energy states. Remarkably, the lower energy spectrum resembles the quasi-hole spectrum in bosonic Moore-Read state at filling fraction $\nu=1$.
\begin{figure}
\includegraphics[width=3.7in]{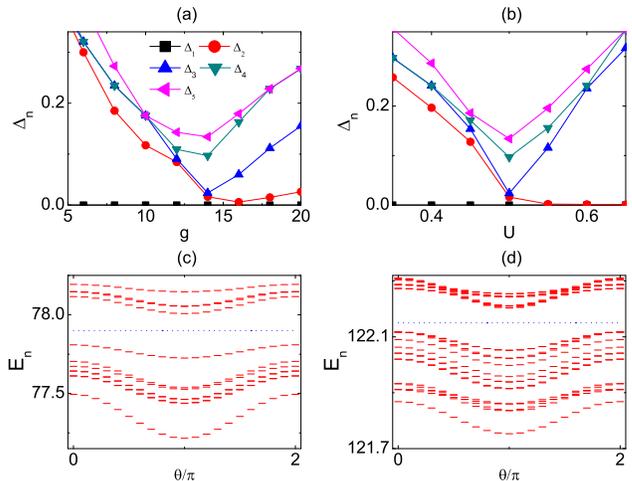}
\caption{(Color online)(a) Excitation energies with respect to the anisotropic parameter $g$ for the spin-1 system with $U=0.5$. (b) Excitation energy with respect to $U$ for the system with $g=14$. (c) Low-energy spectrum with respect to the twist phase $\theta$ for (c) $L=12$ and (d) $L=15$, respectively. Other parameters are $\lambda=0.3$, $\kappa=0.2$, and $\delta=0$.}
\end{figure}

For the continuous system, the $\nu=1$ bosonic Phaffian state in the lowest Landau level \cite{cooper,regnaultstatis} is favored by the three-body repulsive interaction \cite{greiter}. In lattice model, this kind of interaction can be implemented by imposing the three-body hard-core boson constraints \cite{mapping} $(b_i^{\dag})^3=0$, $(b_i)^3=0$. In fact, the spin-$1$ system can be mapped to the system of $3$-body hard-core bosons as they have the same Hilbert space. The mappings between these two systems are: $|0\rangle_{boson}\rightarrow|S_z=-1\rangle$, $|1\rangle_{boson}\rightarrow|S_z=0\rangle$, $|2\rangle_{boson}\rightarrow|S_z=1\rangle$. System with magnetization $m_z= - 2/3$  corresponds to the $\nu=k/(kM+2)$ bosonic Moore-Read state with $k=2$ and $M=0$.
The (2,2)-admissible rules state that any $kM+2$ consecutive orbitals contain at most $k$ particles with the distance between two particles is at least $M$ \cite{Seidel}. The ground state manifold can be $(20),(02),(11)$, which are three-fold degenerate. As the same with spin-$1/2$ case, we have $L/3$ orbitals. For the system with the deviated magnetization $m_z= - 2/3 -1/L$, we only need to count the ways of removing one three-body hard-core boson from the 3-fold ground state manifold. For the system with $L=12$, the following two configurations are allowed: $(2010)$ and $(1110)$ neglecting lattice translation. Similarly, the allowed configurations are $(20200)$, $(11110)$, $(20110)$ for the spin chain with $L=15$. Taking lattice translation into account, finally $8$ and $15$ low energy states should emerge, which agrees with our numerical results.

{\it Summary.-}In summary, we have demonstrated the emergence of fractional quantum states in periodically modulated quantum spin chains with up to third neighbor interactions. By numerical studies based on exact diagonalization and DMRG algorithm, we clearly show the typical character of fractional quantum states: the degenerate ground state manifold, nonzero topological invariants defined in the 2D extended parameter space and the existence of edge states under OBC. The most interesting findings lie in the low energy excitations. For the spin-$1/2$ system, they obey bosonic $\nu=1/2$  fractional statistics. For the spin-1 chain, the statistics resemble those of bosonic $\nu=1$ Moore-Read states. All these evidences clearly demonstrate the existence of fractional quantum states in periodically modulated quantum spin chains. Our research not only makes a link between the periodically modulated quantum spin chain and quantum Hall effect, but also opens an avenue to study fractional quantum states at fractional magnetization in 1D periodic quantum spin systems. As spin chains with tunable interactions have become available in systems of trapped ions or atoms coupled to waveguides, the proposed systems might be a promising platform for exploring exotic fractional statistics.

{\it Acknowledgment.-} We thank Chen Cheng for helpful discussions on the DMRG programs. This work is supported by NSF of China under Grants No. 11425419, No. 11174115, No. 11325417 and No. 11274032.

\end{document}